\documentclass{article}
\usepackage[utf8]{inputenc}
\usepackage[english]{babel}
\usepackage{natbib}
\usepackage{authblk}
\usepackage{booktabs} % For formal tables
\usepackage{enumitem}
\usepackage{graphicx}
\usepackage{url}

\begin{document}

\title{Toward Reducing Crop Spoilage and Increasing Small Farmer Profits in India: a Simultaneous Hardware and Software Solution}
% \titlenote{Produces the permission block, and
%   copyright information}
% \subtitle{Extended Abstract}
% \subtitlenote{The full version of the author's guide is available as
%   \texttt{acmart.pdf} document}

\author[1,2]{George H.~Chen}
\author[2]{Kendall Nowocin}
\author[2]{Niraj Marathe}
\affil[1]{Carnegie Mellon University}
\affil[2]{CoolCrop}

\date{}

% \author{George H.~Chen}
% \affiliation{%
%   \institution{Carnegie Mellon University, CoolCrop}
% }
% % \email{georgechen@cmu.edu}
% 
% \author{Kendall Nowocin}
% \affiliation{%
%   \institution{CoolCrop}
% }
% % \email{kendall@coolcrop.in}
% 
% \author{Niraj Marathe}
% \affiliation{%
%   \institution{CoolCrop}
% }
% \email{niraj@coolcrop.in}

\maketitle

\begin{abstract}
India's agricultural system has been facing a severe problem of crop wastage.
A key contributing factor to this problem is that many small farmers lack
access to reliable cold storage that extends crop shelf-life. To avoid having
leftover crops that spoil, these farmers often sell their crops at unfavorable
low prices. Inevitably, not all crops are sold before spoilage. Even if the
farmers have access to cold storage, the farmers may not know how long to hold
different crops in cold storage for, which hinges on strategizing over when and
where to sell their harvest. In this note, we present progress toward a
simultaneous hardware and software solution that aims to help farmers reduce
crop spoilage and increase their profits. The hardware is a cost-effective
solar-powered
refrigerator and control unit. The software refers to a produce
price forecasting system, for which we have tested a number of machine learning
methods. Note that unlike standard price forecasting tasks such as for
stock market data, the produce price data from
predominantly rural Indian markets have a large amount of missing values. In
developing our two-pronged solution, we are
actively working with farmers at two pilot sites in Karnataka and Odisha.
\end{abstract}

\section{Introduction}

Crop wastage in India results in an annual loss valued at
92,651 crore INR (15B USD) as of 2014~\citep{CIPHET_repeat_2015}.
% This amount increased from
% Rs~44,143 crore (\$6.9 USD) in around 2009.
Among these crops,
fruits and vegetables have the highest wastage percentage, upwards of
15.88\% depending on the crop.
% As high as 15.88 per cent of India's fruit and vegetable harvest is wasted
% annually.
% 
% In India,
% as high as 15.88 per cent of fruit and vegetable harvest is wasted
% annually, contributing to an overall harvest and post-harvest loss valued
% at Rs.~92,651 crore (\$14.4B USD)
% 
% 
% eighteen per cent of India's fruit and vegetable harvest is wasted annually, a
% loss valued at Rs.~92,651 crore (\$14.4B USD) as of
% 2014~\cite{CIPHET_repeat_2015}.
A major part of the
problem is that small and marginal farmers, who as of 2002-2003
account for roughly 81\% of
agriculture holdings in India and who typically have field sizes under 1~hectare
(roughly 2.47 acres)~\citep{dev_2012}, lack access to the required cold storage
% (refrigerators and freezers)
and marketing infrastructure. These
farmers often rely on the cultivation of perishables. Due to lack of access to
cold storage, they are forced to ``crash sell'' their harvest at market prices
dictated by the middlemen or wholesalers to avoid wastage and financial loss.
As an example, a farmer who harvests
100 kg of tomatoes will try to sell them at the nearest market, as soon as
possible, and has to take whatever price is available; otherwise, the produce
spoils and is worth nothing.
% As an example, a farmer who harvests 100 kg of tomatoes will try to sell all of
% them at a nearby market, as soon as possible and at a low price.
With access to
reliable cold storage, the farmer could keep the tomatoes fresh for longer
before selling them at a more favorable price. Reducing food wastage could thus
also increase farmers' profits. With only 10-11\% of fruits and vegetables
produced with access to cold storage, the director of India's National
Horticulture Board stated that a 40\% increase in cold storage capacity would
be needed to avoid wastage~\citep{emerson_2013}.

We are working with small farmers to help them
store and plan when and where to sell their produce. To do so, we provide a
solution that simultaneously has both hardware and software components.
% We seek to address these issues by providing a solution that simultaneously
% has both hardware and software components.
% both hardware and software
% solutions to help small and marginal farmers both store and market their
% products to create more value.
On the hardware side, we are developing
cost-effective efficient solar-powered cold storage units, each
of which is essentially a walk-in closet-like refrigerator that can service
40-50 small farmers. % who band together as a co-op.
On
the software side, we are developing produce price forecasting models with the
goal of helping farmers better plan when and where to sell their produce,
and eventually what and when to grow.
The hardware and software solutions complement each other: without cold storage, the
software solution would not be of much use since a farmer cannot easily delay
selling produce in case of spoilage. With only the cold storage hardware but
not the software solution, it is not straightforward when and where to sell,
and at what price.

We remark that on the software side, the problem we are addressing differs
substantially from, say, forecasting stock market prices or developing a
high-frequency trading strategy. As we discuss in more detail in
Section~\ref{sec:software}, produce
pricing data available for the Indian markets
have a large amount of missing values. To handle these missing values, we
% We remark that on the software side, the problem we are addressing is very
% different from, say, forecasting stock market prices or developing a
% high-frequency trading strategy. In particular, the produce pricing data
% available for the Indian markets (predominantly in rural areas) is substantially
% sparser:
% each day, we collect at most one measurement of a price range (min, mode, max)
% per produce per market, and the availability of historical pricing data is
% highly variable across markets. For example, for certain markets, we have
% pricing information for over five years, whereas for others, we have no historical
% pricing data whatsoever.
% % For any given day, typically at least one
% % market does not have pricing data available.
% Even if we restrict to markets where pricing data are available, many markets
% have less than two years of pricing information.
use ideas from a
forecasting method for clinical time series data that also exhibits a
large data missingness problem~\citep{lipton_et_al_2016}.
Separately,
% Separate from there being exorbitant amounts of missing data,
unlike in normal stock trading or the case of
high-frequency trading,
executing a ``trade'' (i.e., selling some amount of produce at a specific market)
is not remotely instantaneous.
A farmer would have to task someone to drive
to a specific market and stay there for some time to sell produce, easily taking
on the order of hours. Commonly,
farmers choose to sell at multiple markets, which could take more than a whole
day.
% there can easily take multiple hours, and a trip to many markets to sell produce
% could take a whole day.
% the amount of time it takes to drive to all the nearby markets and sell produce
% can easily exceed over 10 hours, just for the driving.
Because of the labor
and time intensiveness of selling, the problem we are tackling could perhaps be
more aptly described by
``very low-frequency trading''. While we focus the discussion of the software
component in this note
only on forecasting and not on the actual execution of ``trades'' (e.g., planning
the driving route to different markets, how long to
stay at each, etc), the latter
clearly suggests that price forecasts should be as far in advance as
possible.

Importantly, we are developing our hardware and software components with input
from local farmers at two pilot sites, one in Dandeli, Karnataka and another in Cuttack, Odisha. Involving local farmers in
agricultural development rather than only providing them with a technological
solution is important to creating a solution that lasts~\citep{dlab_2017}.
We want to ensure that the farmers find our solution to be useful, and we want them to
let us know what we could do better.

% Fishing is quite different from growing produce to
% sell, with typically a much longer planning and upkeep time needed for the latter.
% Thus, even if providing pricing information is not effective for fishermen,
% we suspect that it would be effective for farmers, who may be more inclined to incorporate pricing information into
% their already extensive planning.
% In working with local farmers, our goal in
% providing them with pricing information is to help
% them sell their produce before spoilage.
% To assist

In this note, we report our progress in developing our simultaneous
hardware and software solution. In Section~\ref{sec:hardware}, we describe
our solar-powered cold storage unit.
In Section~\ref{sec:software}, we describe how we cast produce price forecasting
as a classification problem, and benchmark a number of standard classifiers. % as well as some findings on market analytics.
% In Section~\ref{sec:financing}, we discuss how our solution plays into
% the financial context of farmers. In particular, the hardware presents a
% large upfront capital expense. We explain some strategies for securing
% funding for farmers.
We conclude in Section~\ref{sec:future} with a discussion of
end-user financing and
 % of
 future work.
% how our
% solution plays into the financial context of farmers, along with
% future work.

% In this note, we present our current cold storage hardware and produce price
% forecasting method.

% Since starting this project
% early 2016, our initial focus has been on setting up partnerships with farmer
% cooperatives and piloting hardware cold storage solutions, for which we have
% four pilots underway in the Indian states of Bihar, Gujarat, Karnataka, and
% Odisha. We have not started work on the market analytics side.

\section{Cold Storage Hardware}
\label{sec:hardware}

\begin{figure}
\includegraphics[width=.95\linewidth]{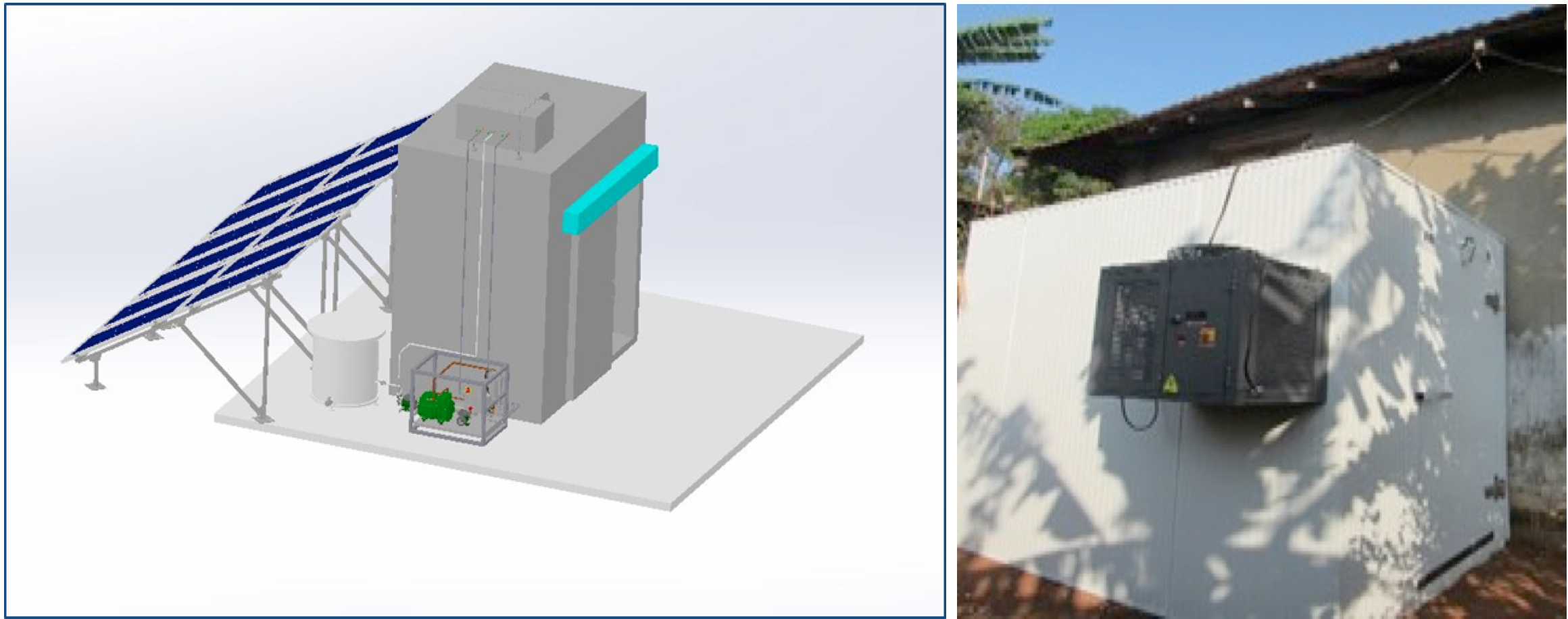}
\vspace{-.5em}
\caption{The
cold storage system: (left) AutoCAD design, (right) deployed unit.}
\label{fig:closet}
\end{figure}

Currently, commercial off-the-shelf (COTS) cold storage units generally consist of an insulated cold room, a cooling unit, and a basic controller. However, for such units, the cooling system generally is incompatible with solar panel systems and consumes a large amount of energy (and is thus not cost-effective). To address these two shortcomings and specifically to tailor the cooling system to the farmers' cultivation behavior and local climate, we developed a new controller, called the \emph{CoolCrop controller}, that readily handles different sensors, cooling units, and power sources including solar. This controller replaces the existing controller of a COTS cold storage unit.
Figure~\ref{fig:closet} shows the CAD and the system deployed in the pilot site of Dandeli, for which we installed our controller in a 5-metric-ton COTS cold storage unit that services about 40-50 small farmers.

\begin{figure}
\includegraphics[width=\linewidth]{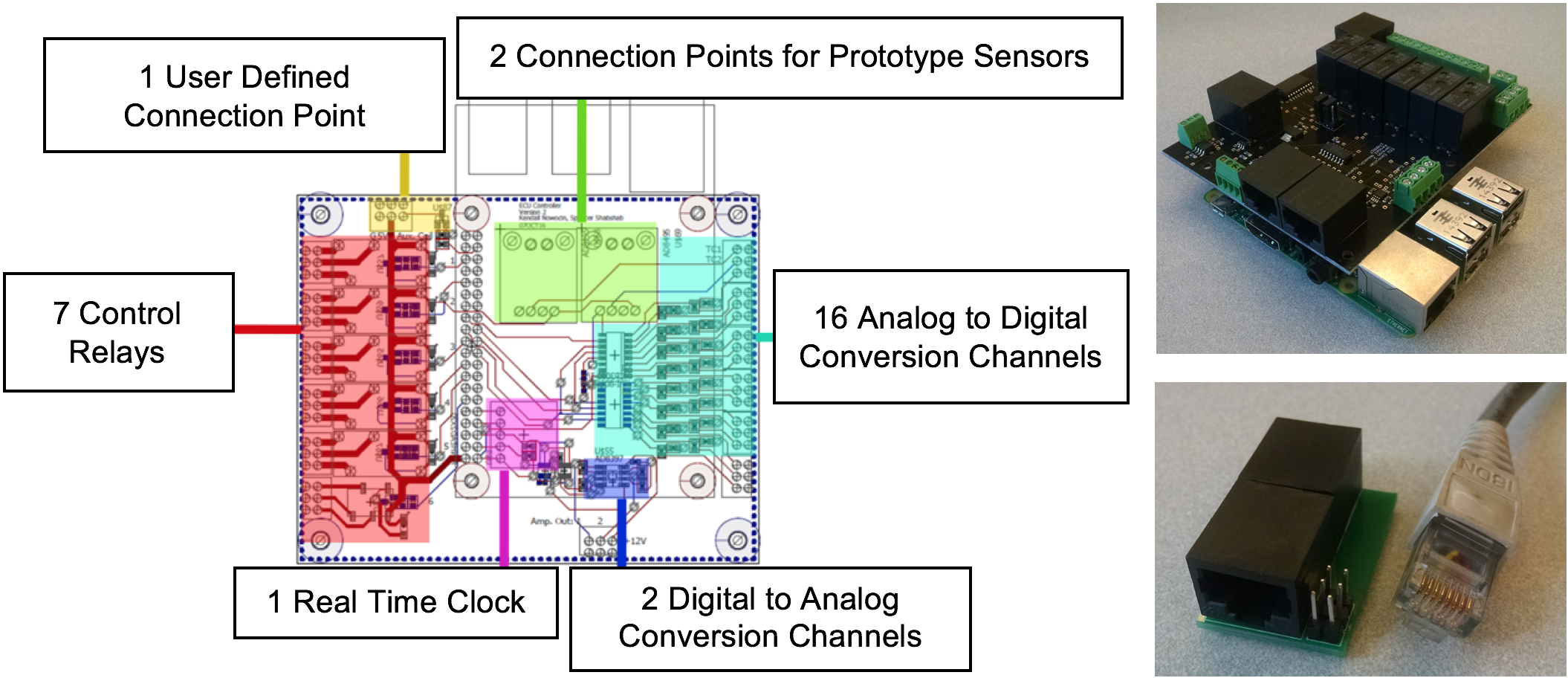}
\caption{The CoolCrop Controller is a low cost data acquisition, control, and processing unit that can be adapted to many applications. Left: controller diagram. Top right: photo of controller. Bottom right: photo of temperature and humidity sensor that connects to the controller via Ethernet.}
\label{fig:controller-schematic}
\end{figure}

The CoolCrop controller records and regulates temperature and humidity within the cold storage unit. We intentionally designed the controller to be flexible in what hardware it can control, what sensors it can pull data from, and what power source it uses. Moreover, we wanted the control logic itself to be easily programmable.
% To both understand how farmers use the cold storage and to minimize overall energy usage of the system, the controller logs temperature, humidity, and power consumption data.
% The new controller board is a
% % We developed a
% low cost data acquisition, control, and processing board for the cold storage unit, which we call the \textit{CoolCrop Controller}.
With these design goals in mind, our controller
consists of a single board Raspberry Pi~3 computer connected to a control and data acquisition board. A diagram and photo of this board is shown in Figure~\ref{fig:controller-schematic}. Low cost sensors for monitoring (i.e., temperature and humidity sensor) can be easily connected using a standard Ethernet cable. The CoolCrop controller has a small form factor approximately 3 inches by 5 inches. % which was useful when integrated into existing equipment and installed as a standalone unit.
Descriptions of the controller functional blocks in Figure~\ref{fig:controller-schematic} are given below:
\begin{itemize}[noitemsep,nolistsep,leftmargin=*]
\item Control Relays: These relays manipulate (energize or de-energize) external control circuits and send alarm flags.
\item User Defined Connection Point: This interface point on the control board can be configured to multiple instrumentation suites.
\item Connection Point for Prototype Sensors: An I2C and SPI interface to connect to and communicate with COTS and integrated sensors. % These are broken out to easy connect the sensor to the acquisition and processing unit. The ports can be multiplex to attached additional sensors.
\item Analog-Digital Conversion (ADC) Channels: These channels can be used to monitor additional aspects of interest and interface with other sensor(s).
\item Digital-Analog Conversion (DAC) Channels: The DACs are used to generate control signals that can be used to achieve more complete control or generate reference or other signals.
\item Real-Time Clock (RTC): The real-time clock is used to generate accurate timestamps for collected data from local, connected, and remote sensors.
\end{itemize}
COTS and custom sensors can easily be integrated to the controller. % COTS temperature sensors were originally used to control the cold storage unit and collect local weather data. The sensors' accuracy ($\pm2^{\circ}$C), resolution (6 bit), and capability (only temperature) were limited, so
We specifically use a new sensor that we have developed that measures temperature within $\pm0.3^{\circ}$C and relative humidity within $\pm$2\%, has user selectable 12 or 16 bit resolution, and connects by a standard Ethernet cable. This sensor (including its circuitry and board)
is approximately the size of two Ethernet jacks; a photo of it next to a Ethernet cable is in the bottom right of Figure~\ref{fig:controller-schematic}.
%   therefore the CoolCrop temperature and humidity sensor board was developed. The bottom right of Figure~\ref{fig:controller} shows the sensor and is accurate on temperature within $\pm0.3^{\circ}$C and relative humidity within $\pm$2\%, user selectable 12 or 16 bit resolution, and connects by a standard Ethernet cable.
% We used it
%  % This was used
%  to demonstrate the easy install, configuration, and calibration of a new sensor connected to the unit. The footprint of the sensor, circuitry, and board is approximately the size of two Ethernet jacks.

\section{Forecasting Produce Prices}
\label{sec:software}

We now discuss produce prices at different markets, and how we forecast these
prices. We collect pricing data from a website called Agmarknet that is run by
the Indian government's
Ministry of Agriculture and Farmers Welfare.\footnote{
\path{http://agmarknet.gov.in/}}
The availability of pricing data varies substantially across markets.
There are markets that have data (price and volume) for over
five
years and others that have no data available.
%  Some
% markets have data available for over 5 years. Otherwise have no data available.
Even for markets with more data available, there are many days for which no
pricing is available for a specific produce, which could be because the produce
is, for instance, out of season, or the data just
was not
recorded.

\begin{figure}
\includegraphics[width=.95\linewidth]{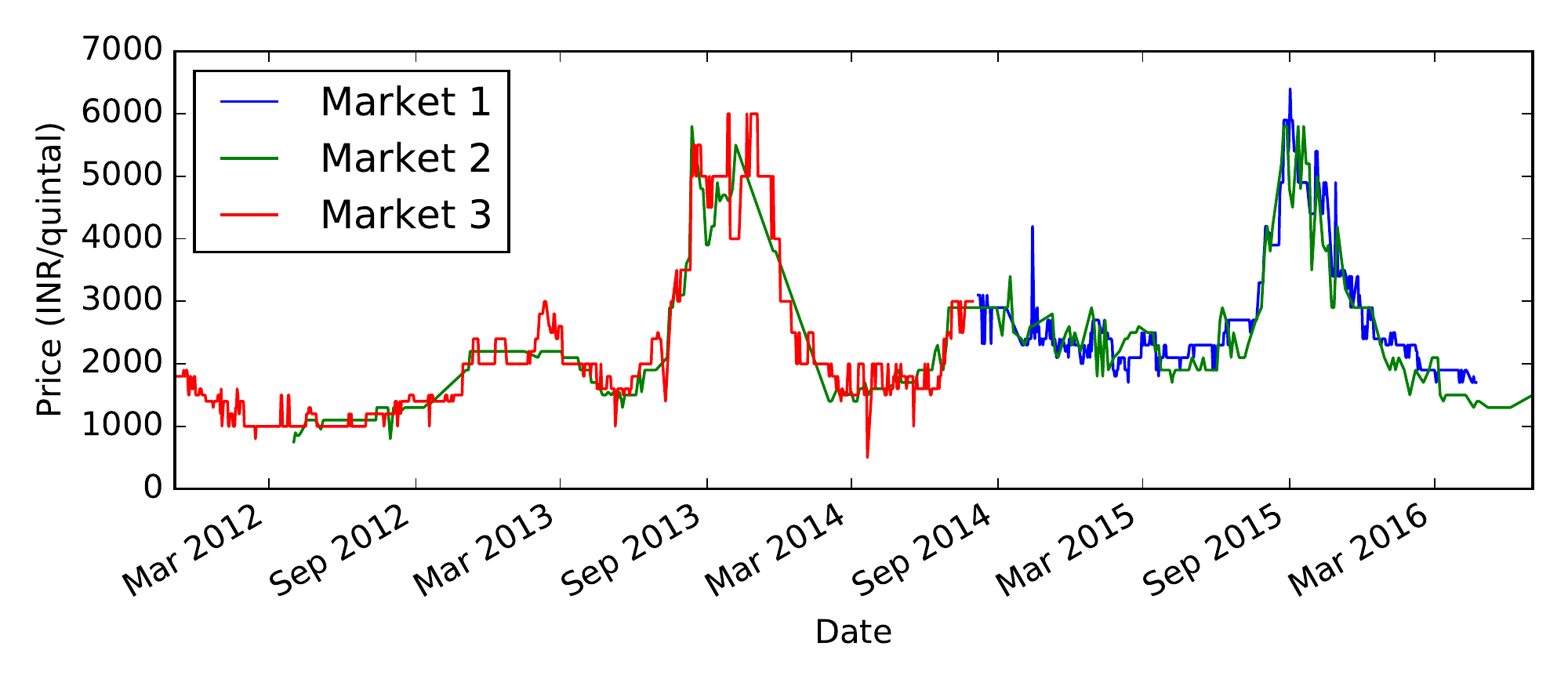}
\vspace{-1.5em}
\caption{Onion prices at three markets in Odisha.}
\label{fig:example-prices}
\end{figure}

An example of onion prices over time is shown in Figure~\ref{fig:example-prices}
for three markets near our second pilot site in Cuttack, Odisha. For these
markets, we collected all available data from 2012 to 2016 off of
Agmarknet. As shown,
onion pricing for the first market
only becomes available in August 2014,
 whereas in the case of the third market, the
pricing information stops being available after July 2014. Within the range of
dates for which pricing data are available, the three markets exhibit
drastically different fractions of missing data. For example, between the
years 2012 and 2016, the first
market only has data from August 7, 2014 to April 22, 2016, where 13.6\% of
the prices are missing between these dates.
For the second market, despite pricing data being
available from
April 1, 2012 up through July 2, 2016, 81.0\% of the days in between do not have pricing
data. % For the third market, 6.9\% of prices are missing. % between
% November 11, 2011, and July 31, 2014.
Thus, even though a market may have pricing data for more years, the data
could be less regularly collected.
% data may be available for a market for more years, it can be
% less regularly collected when it is available.
Lastly,
note that the data exhibit seasonality:
each year between August through March, the price reaches a local
peak.

Forecasting exact prices for the next few days per market turns out to be
challenging. Standard approaches like exponential
smoothing methods~\citep{brown_1956} and
ARIMA models~\citep{asteriou_hall_2011} do not handle missing data well.
% do not handle large amounts of
% missing data well.
Moreover, typically the price of a specific
produce at a specific market does not change over short periods of time,
and predicting the
price at the next day to be the price at the current day is correct
over 60\% of the time in the dataset we collected.
Rather than forecasting exact prices,
we instead forecast just the short-term direction of price changes,
i.e., whether the price will go up, go down, or stay the same for each of
the next few days. Importantly, always predicting that the price stays the
same provides no actionable insight to farmers.

We now describe how to set up a classification problem to
account for missing data and seasonality in forecasting price
movement directions for different markets for a specific produce (onion).
The way we do this is similar to how missing data are modeled in an existing
% We build on an existing
recurrent neural network (RNN) forecasting approach that has been successfully
applied to clinical time series
that contain a large number of missing values~\citep{lipton_et_al_2016}.
Note that the approach of~\citet{lipton_et_al_2016} works for any classifier,
not just RNN's. We similarly will not limit ourselves to only using RNN's. However,
our work differs from that of~\citet{lipton_et_al_2016} in two important ways.
First, we are forecasting price movement directions of the next few days per
time series (each time series is associated with a market). These price movement directions in
general vary with time. In contrast, \citet{lipton_et_al_2016} forecasts a
single non-time-varying outcome per time series. A second difference is that
we explicitly account for seasonality.
% In~\cite{lipton_et_al_2016}, each time series is associated with a single
% outcome to be forecasted, whereas in our setup, we want to forecast future
% values of each time series. Furthermore, our setup has seasonality, which is
% not incorporated in~\cite{lipton_et_al_2016}.
% This approach specifies how to set up a supervised classification problem to
% account for missing data, in fact, works with any classifier
% rather than only with RNN's.

\begin{figure}
\includegraphics[width=\linewidth]{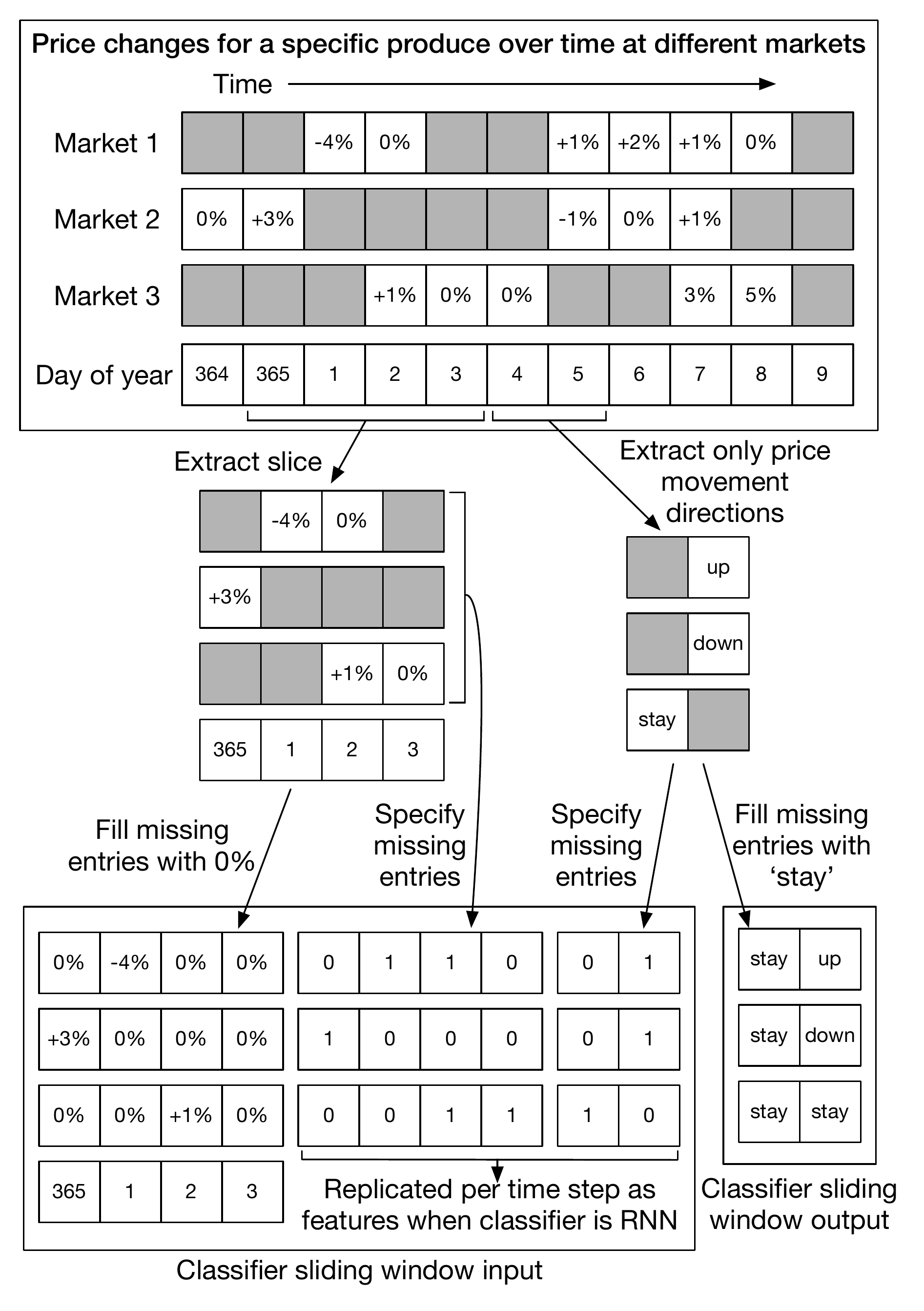}
\vspace{-2em}
\caption{Illustration of how we turn produce pricing data from different
markets into inputs and outputs of a classifier
that, given the last $b=4$ days' price information forecasts the price
movement directions for the next $f=2$ days.
Missing entries are depicted as shaded boxes.} % The days within a year
% wrap around at 366 (to account for leap years), e.g., day~1 corresponds
% to January~1.}
\label{fig:RNN-input-output}
\end{figure}

As shown in Figure~\ref{fig:RNN-input-output}, for a specific produce,
we track percentage price changes of the produce at different markets over
time. Suppose that we want to use
pricing information from the previous $b$ days to forecast price movement
directions
of the next $f$ days. Then to assemble training data, we take a sliding
window approach, each time looking at percentage price changes of the $b$ days
leading up to the current day to treat as input, and the following $f$ days'
price movement directions to use as test data, with a notable exception: we reveal
which entries in the test data are missing as part of training data. The
reason we do this is that we actually only care about predicting entries that
are not missing, which vary over time. Missing data for percentage price
changes in the most recent $b$ days are filled in with 0, while missing
data in price movement directions of the next $f$ days are filled in with
``stay''. Finally, to account for seasonality, we also provide, as additional
inputs to the classifier, the days
in the year that a time window corresponds to (January~1 regardless of
year would be encoded as day~1, January~2 as day~2, etc).
See Figure~\ref{fig:RNN-input-output} for an example where
$b=4$ and $f=2$. % The setup described thus far can be used with a variety
% of classifiers.
Specifically when the classifier used is an RNN, we treat the ``masks'' that
specify which entries are missing as features that we replicate for each
of the $b$ time steps (in the example
of Figure~\ref{fig:RNN-input-output}, this replication procedure would
introduce $6\times3$ features for each of the $b=4$ time steps).
After training the classifier, then to actually do forecasts, we would
provide the classifier with two main inputs:
the percentage price changes of the
most recent $b$ days, and a mask of all 1's specifying that none of the
price movement
directions that we want to forecast are missing.

% We remark that
% our approach differs from that of~\cite{lipton_et_al_2016}
% in two important ways. First,
% in our setup
% we forecast future price movement directions for the next few time
% steps, whereas in~\cite{lipton_et_al_2016}, the goal is to forecast a single
% non-time-varying
% outcome per time series. Second, in our setup, we
% explicitly account for seasonality.

\begin{figure}
\includegraphics[width=.95\linewidth]{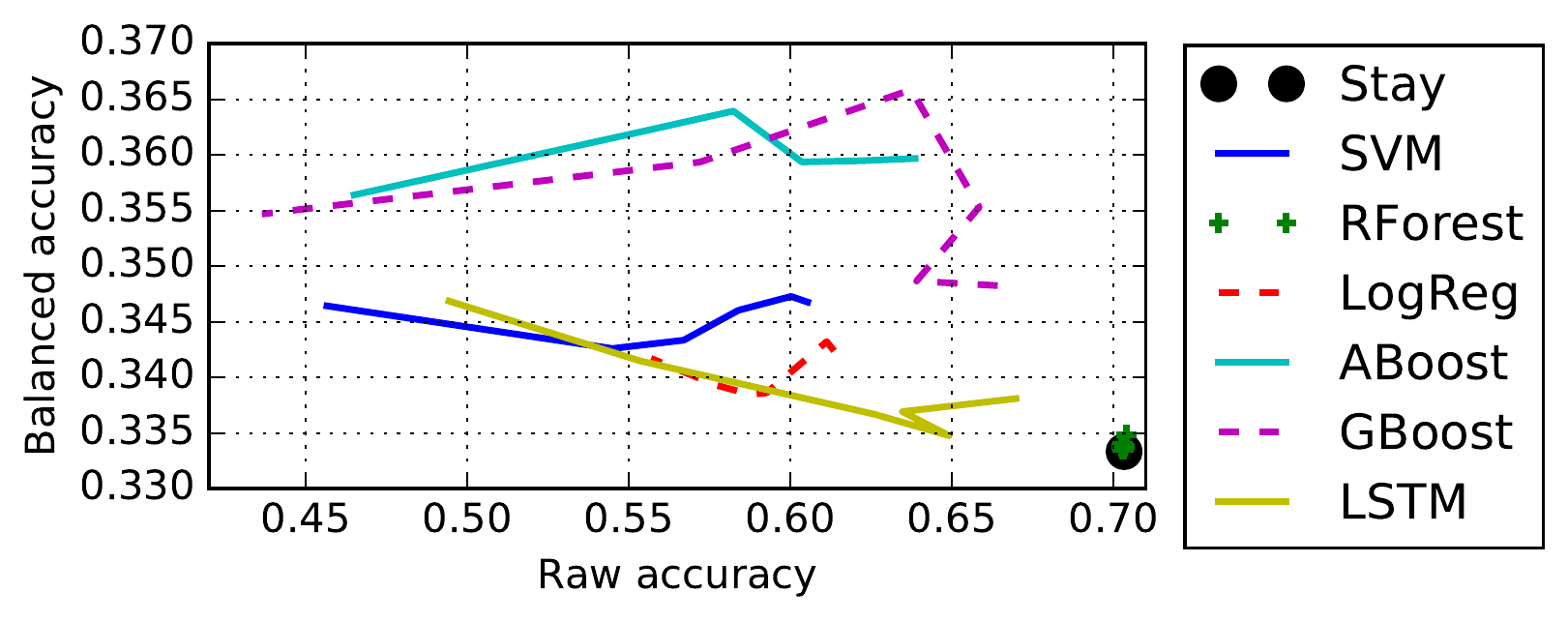}
\vspace{-1em}
\caption{Forecasting results for 2016 produce prices.
% For each method, we vary a parameter $\alpha\in[0,1]$ that specifies how
% much to weight raw accuracy ($\alpha=0$) vs.~balanced accuracy during training.
For each method's curve, moving rightward corresponds to
going from $\alpha=1$ (favor high balanced accuracy) to $\alpha=0$
(favor high raw accuracy).}
\label{fig:test-plot}
\end{figure}

We apply our approach to forecasting onion price movement directions
in a dataset of 14 markets around Cuttack, Odisha. We compared seven different
classifiers for making predictions:
a baseline classifier that always predict that the price will stay
the same (this method is denoted as ``Stay'' in Figure~\ref{fig:test-plot}),
support vector machines (SVM)~\citep{svm}, logistic
regression (``LogReg'')~\citep{logreg}, random forests
(``RForest'')~\citep{random_forests},
AdaBoost (``ABoost'')~\citep{adaboost} with decision trees as base predictors,
gradient tree boosting (``GBoost'')~\citep{grad_boost}, and
long short-term memory (LSTM) RNN's~\citep{lstm_1997}.
We fix the number of days we forecast ahead to
$f=7$. We report two different accuracy measures: (1) a raw accuracy
for what fraction of price movement directions are correctly predicted,
and (2)~the average of three accuracy fractions corresponding to
correctly predicting price movement directions
going up, going down, and staying the same (we call this the
``balanced'' accuracy). The balanced accuracy measure
helps deal with class imbalance between the three outcomes, especially
as over 60\% of the time the price stays the same. Note that asking for
high raw accuracy differs from asking for high balanced accuracy.
We introduce a parameter $\alpha\in[0,1]$ for the user to choose that
specifies how important balanced accuracy is when training the classifier
($\alpha=0$ means we only care about raw accuracy, and
$\alpha=1$ means we only care about balanced accuracy).
We train on data
from 2012 through 2015 and forecast price movement directions in 2016. We tune classifier
parameters (but not $\alpha$, which the user specifies)
during training by treating 2015 pricing information as validation data.
% We ask the user to specify a parameter $\alpha\in[0,1]$ that controls
% how much to weight raw vs.~balanced accuracy ($\alpha=1$ corresponds to
% only caring about balanced accuracy).
% One parameter that we ask the user specify is how much to weight raw
% accuracy vs.~balanced accuracy.
% 
% One parameter that
% One parameter that we leave up to the user to specify is
% $\alpha\in[0,1]$, % that we let the user choose,
% which specifies how much to weight raw accuracy vs.~balanced accuracy
% during training;
% $\alpha=0$ corresponds to only caring about
% raw accuracy, and $\alpha=1$ corresponds to only caring about balanced
% accuracy.
% tuning parameters is done via holding out part of the training data as a
% validation set (specifically we use year 2015 pricing information for
% validation).
% We use the most recent
% $b=14$ days to forecast (we found $b=7$ to often give similar results,
% which were better than $b=1$).

Raw and balanced forecasting accuracies
on the 2016 test data for
varying $\alpha$ are shown in Figure~\ref{fig:test-plot}; note that for
each method, we average over all $f=7$ days and all 14 markets in computing
the two accuracy measures. For all methods, we forecast based on
pricing information for the most recent
$b=7$ days ($b$ is chosen during training using validation data). We see that the
trivial Stay classifier has the best raw accuracy but the worst balanced
accuracy. The random forest classifier effectively learns to nearly
almost predict ``stay'' and thus performs similar to the Stay
classifier. Meanwhile, gradient tree boosting and AdaBoost
achieve the best tradeoff between raw and balanced accuracies.
For example, gradient tree boosting with $\alpha=0.6$ correctly predicts
20.4\% of up
movements, 15.7\% of down movements, and 73.6\% of stay movements
to achieve a raw accuracy of 63.7\% and balanced accuracy of 36.6\%.
We remark that since gradient tree boosting, AdaBoost with decision tree
base predictors, and random forests are adaptive nearest neighbor methods,
% These two methods
% also have
they
can
provide evidence for their forcasts in the form of past training data most
similar to test
data. This evidence may be helpful to farmers. % Farmers may find such evidence helpful.
% which can help farmers interpret the forecasts.

%  Raw and
% balanced accuracies of the forecasts are shown in
% Figures~\ref{fig:forecast-results-raw}
% and~\ref{fig:forecast-results-balanced}.
% The
% random forest (RForest) learns to almost always predict ``stay''
% and thus has performance very similar to the baseline trivial
% ``Stay'' forecaster. Since by far the most common outcome for price
% changes is to stay, unsurprisingly these two methods achieve the
% highest raw forecast accuracy. However, they have among the worst
% balanced forecasting accuracy as they effectively do not learn a
% good model for predicting when there is a price change.
% % which readily achieves high accuracy (due to ``stay'' being
% % by far the most common outcome) but does not produce a useful
% % model for when the price changes, which is of interest to farmers.
% The method that simultaneously is able to achieve among the highest
% balanced \textit{and} raw forecast accuracies is the LSTM RNN
% forecaster.
% % Even so, the highest balanced forecast accuracies achieved thus far
% % are still low compared to the highest raw accuracies.

% The method that is able to learn more about price changes and that
% has the highest raw accuracy amongst the methods tested is the
% LSTM RNN forecaster.

\section{Discussion}
\label{sec:future}

% TODO: future work on hardware and software (discussion of cold storage for transportation to reduce spoilage mid transportation, also for fishermen)

\textit{End-user financing.} 
One of the challenges in providing new technological solutions such as what
we have developed is end-user financing, especially to cover the capital cost
of hardware. The small farmers we have talked to make on average around 40,000
INR (\$600 USD) per year from agriculture, and are hesitant to invest or take
loans to pay for the system. Many of them not only lack access to financial
products and services, but they also do not have a credit history. This problem can be addressed by
creating low-risk financial models and
streamlining the process of financing through formal routes such as banks,
micro-financing institutes, and government subsidies.
To successfully implement these financial models, we create partnerships with local organizations, NGOs, and farmer cooperatives who can ensure that the farmers use the storage judiciously and stick to payment schedules, which can either be based on renting the hardware for as long as the farmers find it useful, or paying for it over a duration of 3-5 years.

In our pilot site of Dandeli, we have observed that farmers increase their vegetable production by at least 100\% after they have cold storage. Considering the amount of wastage curbed and overall increase in vegetable production, we suspect the annual income of farmers to dramatically increase. In devising postpaid models for small farmers to cover the initial capital cost, we account for both their increase in revenue and their existing financial situation.

\textit{Future work.} On the hardware side,
we are working on making the cold storage unit cheaper and more energy efficient.
% making the cold storage energy efficient is the biggest challenge.
The controller needs to minimize both the surge power consumption during the start-stop mechanism of the cooling equipment, and the energy consumption while maintaining the desired temperature and humidity. We are also looking into cheap, locally-sourced building materials such as thermal storage and additional insulation to both
decrease the system cost and improve its energy efficiency. % such as thermal storage and insulation. % such as thermal storage and various insulation materials to improve energy efficiency.

On the software side, we suspect that forecasts could be improved by accounting for a variety of real-time parameters such as rainfall, market demand, and available supply of produce. However, having good forecasts is not enough. Figuring out how to communicate forecasts to farmers is extremely important. % , and how to gauge uncertainty in the forecasts.
% Understanding how to most effectively present
% pricing information to farmers is important;
% Otherwise they could easily ignore
% forecasts. % and
% use the cold storage hardware without carefully planning
% when and where to sell.
For example, sending SMS messages
about prices at different markets could be insufficient. % In fact, in the context
% of fishing, there is debate as to whether such as SMS strategy
% actually helps
% fishermen.
In a 2007 report, Jensen claimed that such an SMS strategy significantly helps fishermen identify which market to
sell at~\citep{jensen_2007}. However, a recent study by Steyn refutes Jensen's
claims and gives evidence that fishermen do not take advantage of
such information despite its availability~\citep{steyn_2016}. We are working
with farmers to understand what forecast information they will find most useful.
% To avoid this
% scenario, we are gathering input from farmers to understand what they find
% useful.
Moreover, we suspect that since growing and selling horticulture crops
typically requires
more extensive planning than fishing, horticulture farmers may be more inclined to
 take
advantage of pricing information than fishermen.

\vspace{0.5em}
\noindent
\textbf{Acknowledgments.} This work was supported in part by the MIT Legatum Center, MIT Sandbox, and the CMU Berkman Faculty Development Fund.

\bibliography{coolcrop_lft} 

\end{document}